\documentclass[useAMS,usenatbib,usegraphicx]{mn2e}
\usepackage{epsfig}
\usepackage{amsmath} %need for line break in equ 3
\usepackage{rotating}           % for sideways tables/figures
\usepackage{color}     
\usepackage{graphicx}
\usepackage{times}
\usepackage{upgreek} % for up $\uptau$ 
\usepackage{amssymb} % for \checkmark

\def\kms{km ${\rm s}^{-1}$}

\def\ccm {$\hbox{{\rm cm}}^{-3}$}    %cm-3
\def\scm  {$\hbox{{\rm cm}}^{-2}$}    %cm-2
\def \AL {$\alpha $}     %  gr. alpha
\def \HI {H{\sc \,i}}
\def \WpHz {W Hz$^{-1}$}
\def\lapp{\ifmmode\stackrel{<}{_{\sim}}\else$\stackrel{<}{_{\sim}}$\fi}
\def\gapp{\ifmmode\stackrel{>}{_{\sim}}\else$\stackrel{>}{_{\sim}}$\fi}
\title[A Critical Ionising Luminosity in Active Galaxies]{Further Observational Evidence for a Critical Ionising Luminosity in Active Galaxies}
\author[S. J. Curran,  et al.]{S. J. Curran$^{1}$\thanks{Stephen.Curran@vuw.ac.nz},  R. W. Hunstead$^{2}$,  H. M. Johnston$^{2}$, M. T. Whiting$^{3}$, E. M. Sadler$^{2}$,  
\newauthor  J. R. Allison$^{3}$ and  C. Bignell$^{4}$\\
$^{1}$School of Chemical and Physical Sciences, Victoria University of Wellington, PO Box 600, Wellington 6140, New Zealand\\
$^{2}$Sydney Institute for Astronomy, School of Physics, The University of Sydney, NSW 2006, Australia\\
$^{3}$CSIRO Astronomy and Space Science, PO Box 76, Epping NSW 1710, Australia\\
$^{4}$National Radio Astronomy Observatory, P.O. Box 2, Rt. 28/92 Green Bank, WV 24944-0002, USA}

\begin{document}

\date{Accepted ---. Received ---; in original form ---}

\pagerange{\pageref{firstpage}--\pageref{lastpage}} \pubyear{2017}

\maketitle

\label{firstpage}

\begin{abstract}
  We report the results of a survey for \HI\ 21-cm absorption at redshifts of $z\gapp2.6$ in a new sample of radio
  sources with the Green Bank and Giant Metrewave Radio Telescopes. From a total of 25 targets, we report zero
  detections in the 16 for which optical depth limits could be obtained.  Based upon the detection rate for $z\geq0.1$
  associated absorption, we would expect approximately four detections.  Of the 11 which have previously not been
  searched, there is sufficient source-frame optical/ultra-violet photometry to determine the ionising photon rate for
  four. Adding these to the literature, the hypothesis that there is a critical rate of $Q_\text{\HI}\sim10^{56}$
  ionising photons per second (a monochromatic $\lambda=912$ \AA\ luminosity of $L_{\rm UV}\sim10^{23}$ \WpHz) is now
  significant at $\approx7\sigma$. This reaffirms our assertion that searching $z\gapp3$ active galaxies for which
  optical redshifts are available selects sources in which the ultra-violet luminosity is sufficient to ionise all of
  the neutral gas in the host galaxy.
\end{abstract} 


\section{Introduction}
\label{intro}

While absorption of the 21-cm spin-flip transition of neutral hydrogen (\HI) is readily detected in near-by galaxies, it
is conspicuous by its absence from the redshifted Universe. \HI\ 21-cm absorption traces the cool component of the gas,
the reservoir for star formation, which can be detected in either quiescent galaxies {\em intervening} the line-of-sight
to a more distant radio source or {\em associated} with the host galaxy of the radio source itself.  While the detection
of hydrogen at radio wavelengths is not susceptible to the dust extinction effects of optical/ultra-violet band searches
\citep{cmr+98,cwm+06,cwa+17}, after over three decades of searching (e.g. \citealt{dm78,dsm85}) there are only 50
detections each of \HI\ 21-cm at $z\gapp0.1$ in either intervening or associated absorption (compiled in
\citealt{cdda16}).  This is despite \HI\ being detected through Lyman-\AL\ absorption in 12\,000 damped Lyman-\AL\
absorbers (DLAs) and sub-DLAs\footnote{A DLA is defined as having a neutral hydrogen column density exceeding $N_{\rm
    HI} = 2\times10^{20}$ atoms \scm, with sub-DLAs having lower column densities, although exhibiting the same
  characteristic damping wings in the absorption profile.} \citep{npc+12}, mainly at $z\gapp1.7$ where the Lyman-$\alpha$
transition is redshifted into the atmospheric observing window.\footnote{At lower redshifts, a high \HI\ column density
  is inferred from the equivalent width of the Mg{\sc \,ii} absorption and other metal ion transitions which can be
  detected by ground-based telescopes (e.g. \citealt{rtn05}).}  By comparison, the detection of \HI\ 21-cm absorption
occurs overwhelmingly at redshifts of $z\lapp1$, or less than half the look-back time to the origin of the
Universe. While much of this can be attributed to the past availability of the suitable radio bands, detection rates at
$z\gapp1$ are demonstrably lower than at $z\lapp1$ (\citealt{kc02,cwt+12}).

For the intervening absorbers it has been argued that this is due to an evolution in the spin temperature, $T_{\rm s}$,
of the gas, where the fraction of the cold neutral medium (CNM, where $T\sim150$ K and $n\sim10$ \ccm) decreases with
redshift \citep{kc02,kps+14}. However, given that the comparison of the 21-cm absorption strength with the total neutral
hydrogen column density, from the Lyman-\AL\ transition, can at best yield the spin temperature/covering factor
degeneracy ($T_{\rm s}/f$), \citet{cmp+03} argued that an apparent evolution in $T_{\rm s}$ could be caused by the
assumption that $f=1$ for the high redshift absorbers.  
Furthermore, by accounting for the effects of an expanding Universe, a standard $\Lambda$ cosmology means that at
redshifts of $z\gapp1$ the absorber is always at a similar angular diameter distance as the background continuum source,
i.e. $DA_{\rm abs} \approx DA_{\rm QSO}$. In contrast, at $z\lapp1$ $DA_{\rm abs} < DA_{\rm QSO}$ is possible (when
$z_{\rm abs} < z_{\rm QSO}$), suggesting that the mix of ``spin temperatures'' at $z\lapp1$ and exclusively high values
at $z\gapp1$ \citep{kc02}, is in fact the effect of geometry on the covering factor \citep{cw06,cur12}. By
deconstructing the covering factor, via the angular diameter distances and high resolution imaging of the background
sources, there is now compelling evidence that the fraction of the CNM in intervening absorbers may trace the star
formation history of the Universe \citep{cur17}.

For the associated absorbers, since, by definition, $z_{\rm abs} \approx z_{\rm QSO}$, the same geometric effects cannot
account for the difference in the detection rates between the low and high redshift sources. From a survey of \HI\ 21-cm
absorption at $z\gapp3$, \citet{cww+08} suggested that their exclusively non-detections were due to the high redshifts
selecting sources with high rest-frame ultra-violet luminosities ($L_{\rm UV}\gapp10^{23}$ \WpHz). Since an examination
of the literature showed that \HI\ 21-cm had never been detected at these luminosities, irrespective of redshift, they
suggested that the non-detection of cool neutral gas was due to the high redshift selection  yielding only
sources where the UV luminosity was sufficient to excite the gas to below the detection limit.  
This observational result has since been confirmed several times for various heterogeneous samples
(\citealt{cwm+10,cwsb12,cwt+12,caw+16,cwa+17,ace+12,gmmo14,akk16,gdb+15})\footnote{\citet{gdb+15}, which is still in
  submission, reports 0 new detections of \HI\ 21-cm absorption out of 89 new searches over $0.02 < z < 3.8$ (see
  \citealt{gd11}).}, showing this to be an ubiquitous effect.  

In order to investigate why there is an apparent critical
UV luminosity and why this is $L_{\rm UV}\sim10^{23}$ \WpHz, \citet{cw12} applied the equation of photoionsation
equilibrium \citep{ost89} to a gas disk with an exponential density distribution \citep{bbs91} and found that an
ionising photon rate of $Q_\text{\HI}\equiv \int^{\infty}_{\nu}({L_{\nu}}/{h\nu})\,d{\nu}\approx3\times10^{56}$~s$^{-1}$
(a monochromatic $\lambda=912$~\AA\ luminosity of $L_{\rm UV}\sim10^{23}$ \WpHz) will ionise a gas disk with the same
scale-length as the Milky Way \citep{kdkh07}. This suggests that the observed critical luminosity is sufficient to
ionise all of the neutral gas in a large spiral galaxy, thus explaining why \HI\ 21-cm has never been detected where
$L_{\rm UV}\gapp10^{23}$~\WpHz.  Since the vast majority of $z\gapp1$ radio sources for which redshifts are available
are believed to have luminosities above the critical value (see figure 4 of \citealt{msc+15}), this would mean that even
the Square Kilometre Array (SKA) will be unable to detect \HI\ 21-cm absorption in these objects.

Given this critical luminosity above which all of the gas in the host is ionised, in order to detect the cool,
star-forming gas within high redshift radio sources, we need to dispense with the reliance upon an optical redshift,
thus avoiding the selection of the most UV luminous objects.  Through their wide instantaneous bandwidths, the SKA
pathfinders are ideally suited to spanning a large range of redshift space in a single tuning
(e.g. \citealt{asm+15,mmo+17}).\footnote{Since the nature of the absorber is usually determined from an optical
  spectrum, other techniques for distinguishing intervening from associated absorption must be explored, with machine
  learning showing some promise \citep{cdda16}.} However, these are generally limited to frequencies of $\gapp700$ MHz,
i.e. \HI\ 21-cm at redshifts of $z\lapp1$, therefore requiring the SKA to find the missing cold neutral gas at high
redshift.  In the meantime, we can continue to search for \HI\ 21-cm absorption, via various search strategies, with
currently available instruments (e.g. \citealt{caw+16}). From newly obtained redshifts for a sample of strong flat
spectrum radio sources (see Sect. \ref{ss}), we can perform a large survey for associated 21-cm absorption in active
galaxies over redshift ranges spanning most of the Universe's history.  Here we present the results of our high redshift
($z\gapp2.6$) survey with the Green Bank Telescope (GBT) and the Giant Metrewave Radio Telescope (GMRT).

\section{Observations and analysis}
\label{sec:obs}

\subsection{The sample}
\label{ss}

The {\em Second Realization of the International Celestial Reference Frame by Very Long Baseline Interferometry} (ICRF2,
\citealt{mab+09}), constitutes a sample of strong flat spectrum radio sources, of which 1682 now have known redshifts
(\citealt{tm09,tsj+13} and references therein), yielding a tuning frequency in the search for 21-cm absorption.  Being
VLBI sources, all have significant compact flux, thus maximising the chance of a high covering factor and thus optical
depth \citep{cag+13}.  The original aim of the survey was to form part of a large observing campaign to search and
quantify the incidence of associated \HI\ 21-cm absorption over all redshifts, although observing time was only granted
for the high redshift ($z \gapp2.6$) proposals.  As mentioned above (Sect. \ref{intro}), we believe that such a high
redshift selection will yield only sources above the critical UV luminosity, where the detection of \HI\ 21-cm is
unlikely. However, having a preconceived suspicion of the outcome should not preclude it from being tested.

\subsection{Observations and data reduction}
\label{sec:obs}

\subsubsection{GBT observations}

In the ICRF2, 1456 of the sources have declinations accessible to the GBT, in addition to being at a redshift which
places the 21-cm transition into a GBT band. For this, the high redshift survey, we used the PF1 290--395 MHz receiver
(\HI\ 21-cm over $z = 2.6-3.9$).  We formed our target list by prioritising sources with flux densities estimated to be
in excess of 1~Jy at the redshifted 21-cm frequency.  Excluding the two which already had published searches at the
time, left 24 targets.  The observations were performed over 24--26 February 2014 with each source observed for a total
of 1.3 hours in two orthogonal linear polarisations ({\sf XX} \& {\sf YY}). The Prime Focus 1 (PF1) receiver was used,
backed by the GBT spectrometer, with a bandwidth of 12.5 MHz, in order to minimise RFI while maintaining a velocity
coverage of $\Delta v \approx \pm5500$~\kms.  A channel width of 11 kHz gave a spectral resolution of
$\approx10$~\kms. The data were analysed, flagged for RFI and averaged using the {\sc gbtidl} software.

\subsubsection{GMRT observations} 

In the ICRF2, 416 of the sources have declinations accessible to the GMRT, of which 54 have 21-cm shifted to within the
90-cm band (305--360 MHz, \HI\ 21-cm over $z = 2.9-3.7$).  We prioritised sources with flux densities estimated to be in
excess of 0.5~Jy at the redshifted 21-cm frequency.  As per the GBT observations, the flux limit was chosen in order to
attain high sensitivities, while still retaining a sizable number of targets. If the radio luminosity is correlated with
the UV luminosity, it is conceivable that this selection may bias against the detection of cool, neutral gas. Although there
is a correlation \citep{cw10}, there is no critical rest-frame 1.4 GHz luminosity \citep{cww+08}, with the relationship
probably  arising from both luminosities being correlated with the redshift.

Excluding the potential targets with published searches for 21-cm absorption at the time left 16 targets.  The observations were
taken with the full 30 antenna array over 9--13 November 2013 and 28--29 March 2014, with each source observed for a
total of two hours in two orthogonal circular polarisations ({\sf LL} \& {\sf RR}). For bandpass calibration 3C\,48,
3C\,147 and 3C\,298 were used, with the phases being self calibrated apart from 1427+543, 1427+543, 0521--262 \&
2318--087, which used a strong near-by unresolved source.  For the backend we used the FX correlator over a bandwidth of
4 MHz, which over 512 channels gave a spectral resolution of $\approx7$ \kms. The data were calibrated and flagged using
the {\sc miriad} interferometry reduction package.  After averaging the two polarisations, a spectrum was extracted from
the cube. None of the sources was resolved by the synthesised beam, which ranged from $10.7\arcsec\times9.4\arcsec$ to
$30.2\arcsec\times11.1\arcsec$.
 
\section{Results}
\label{sec:res}
\subsection{Observational results} 

\begin{figure*} 
\vspace{18.7cm}  
\includegraphics{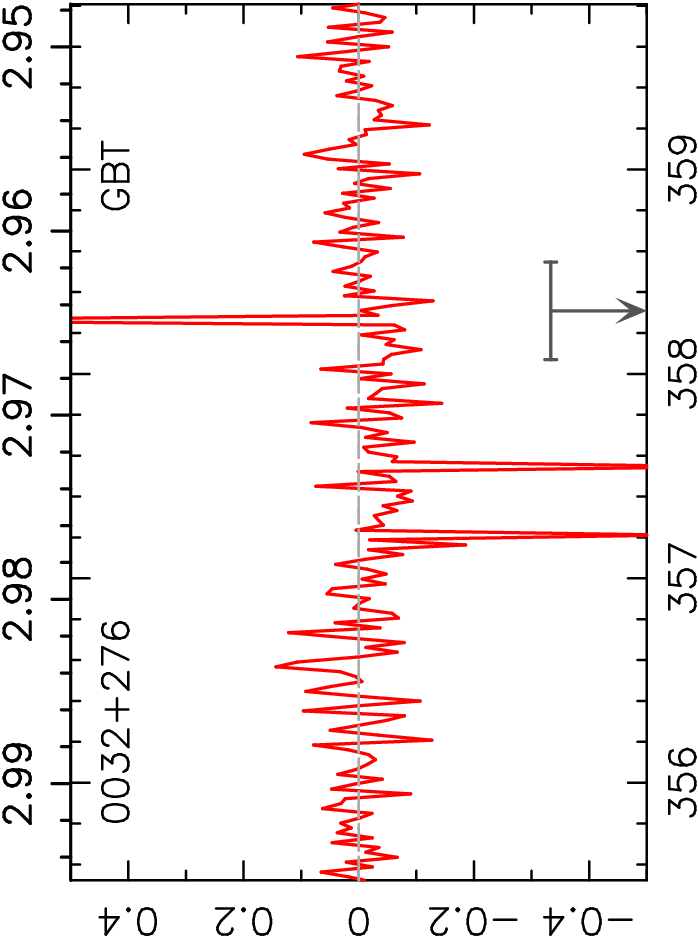}  
\includegraphics{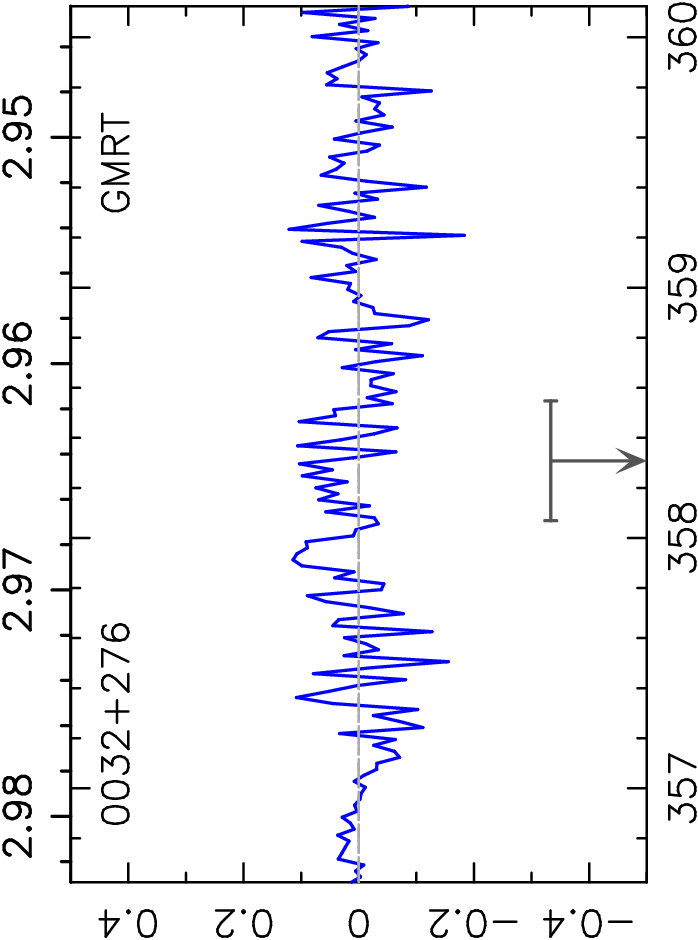} 
\includegraphics{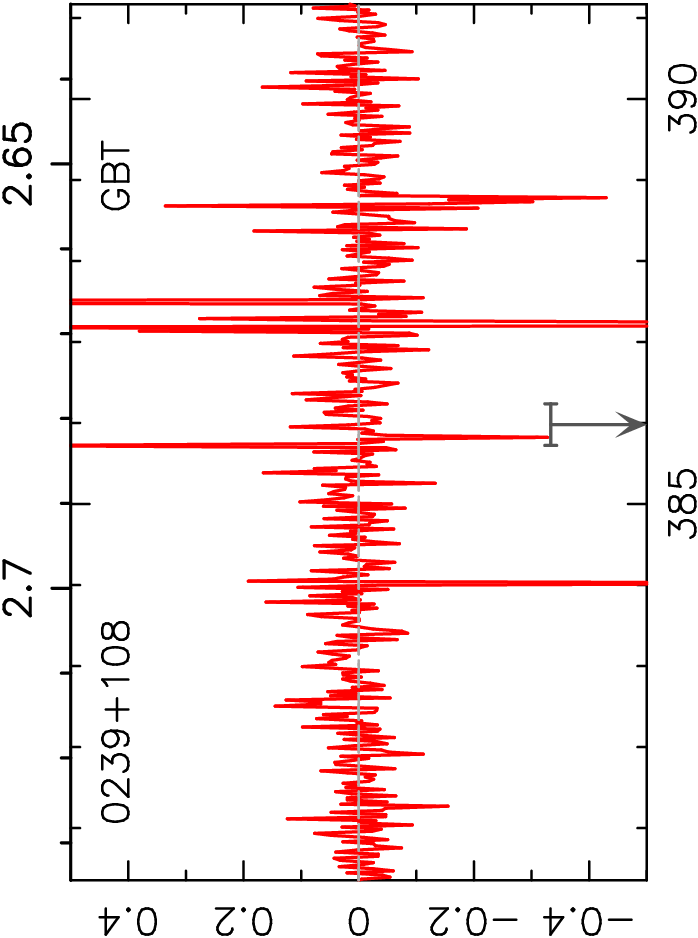} 
\includegraphics{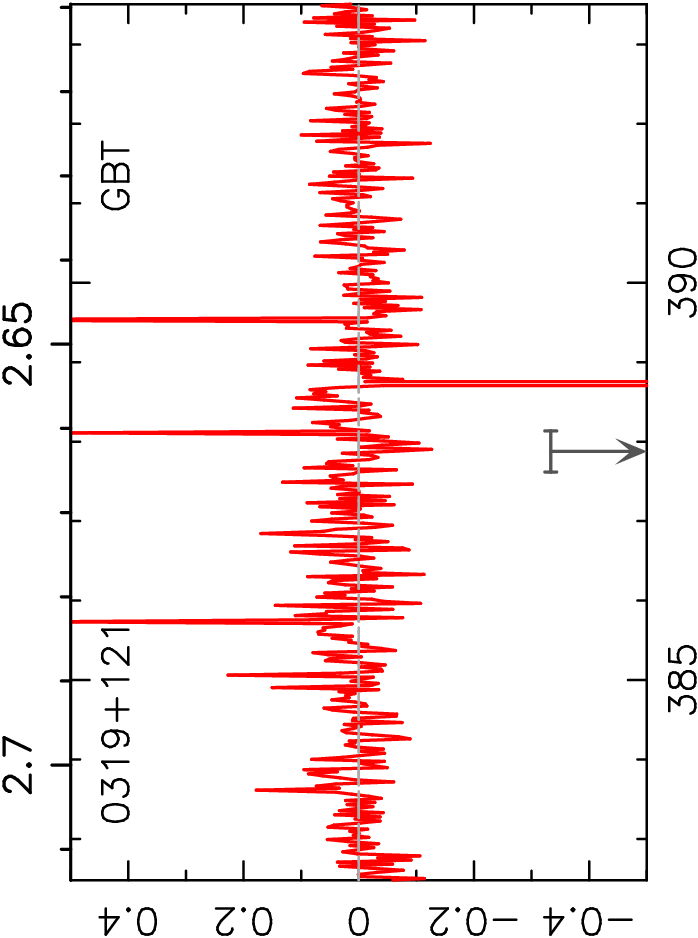}
\includegraphics{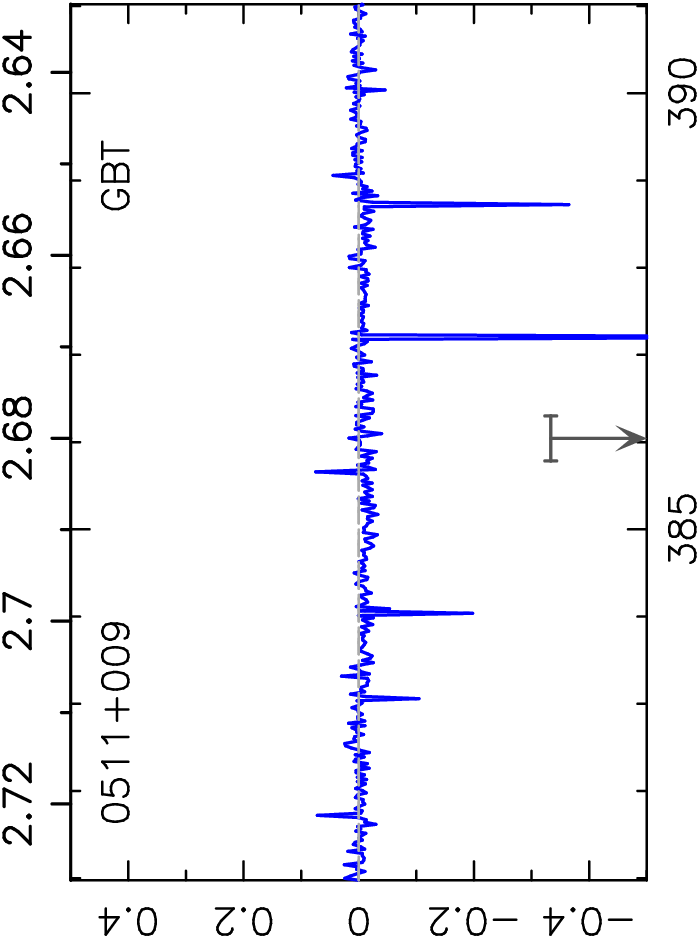}  
\includegraphics{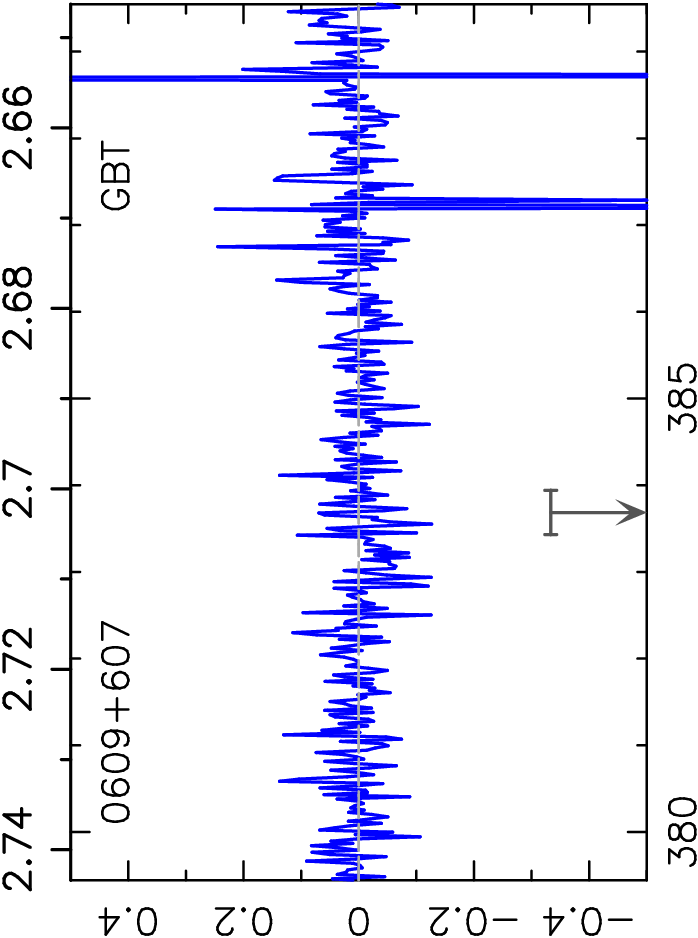} 
\includegraphics{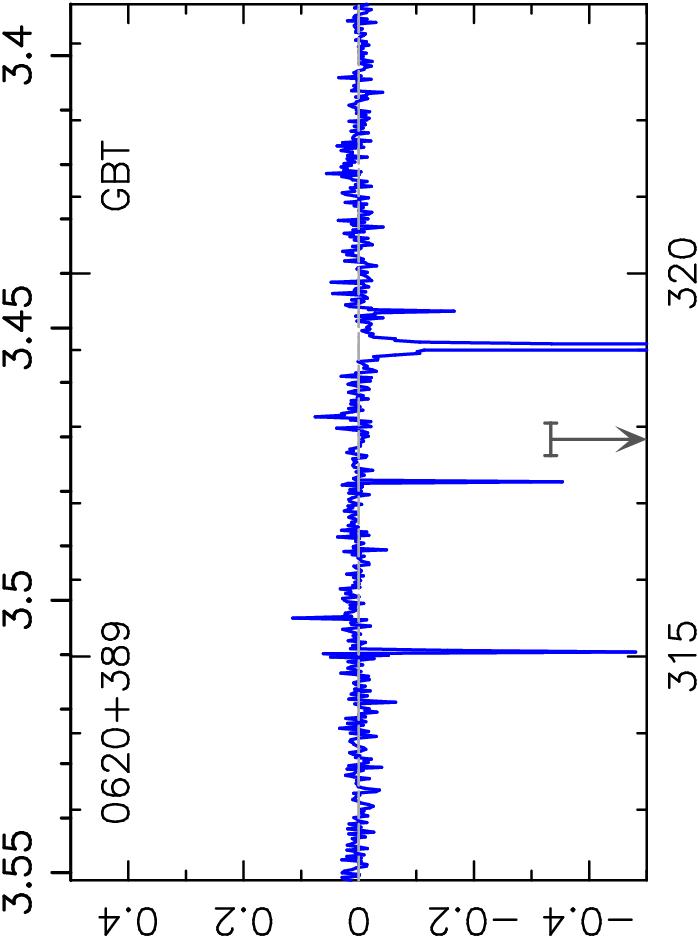} 
\includegraphics{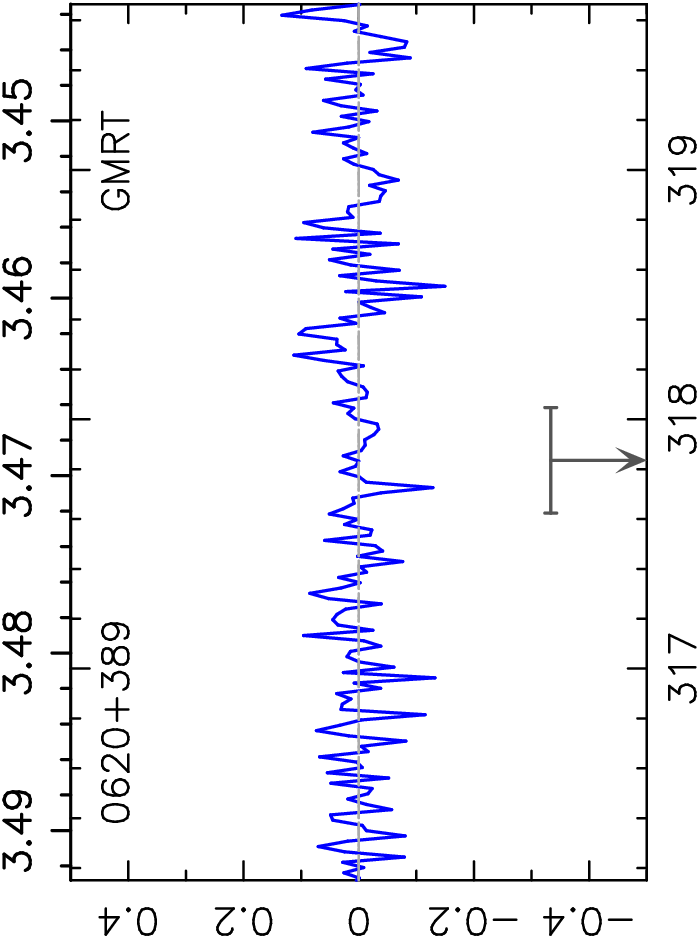}
\includegraphics{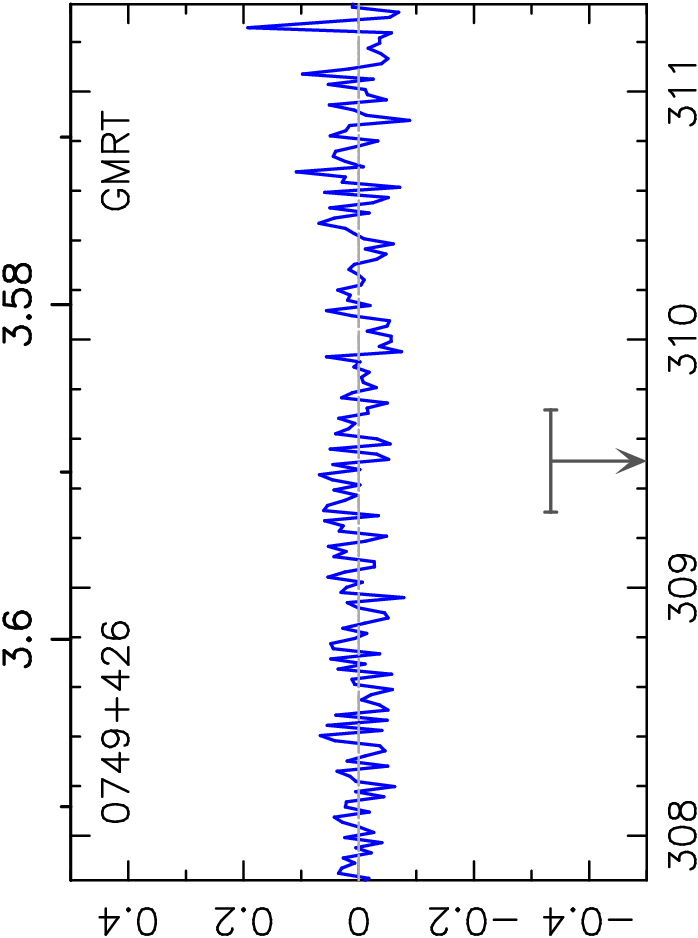}  
\includegraphics{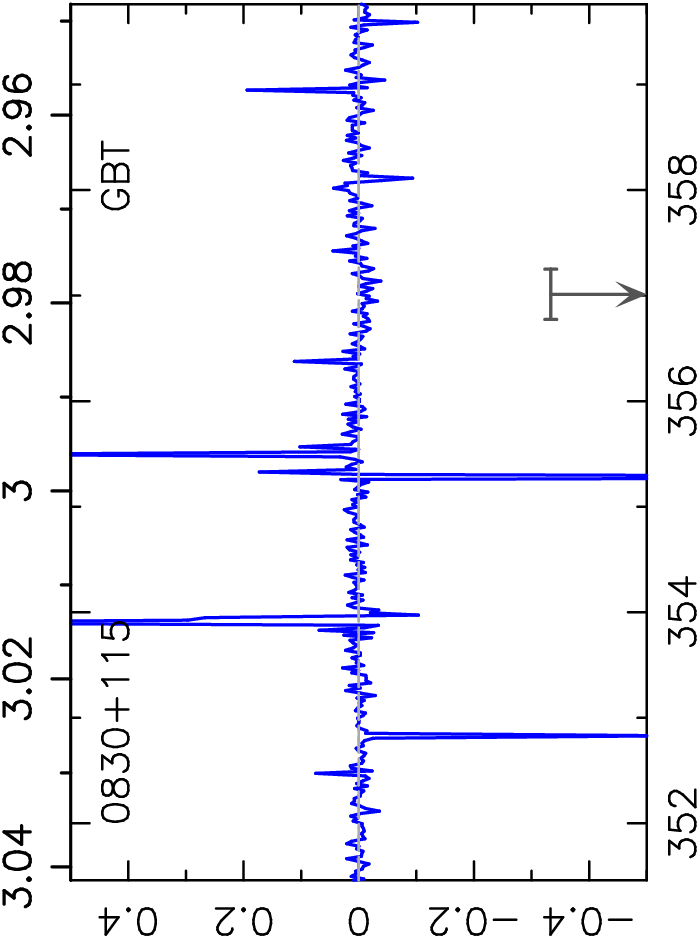} 
\includegraphics{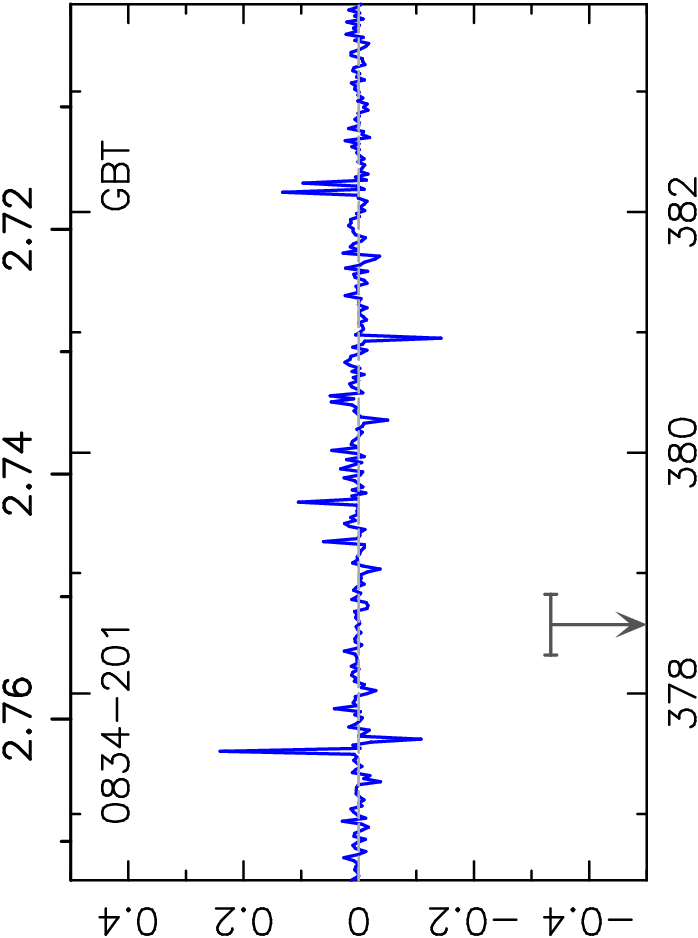} 
\includegraphics{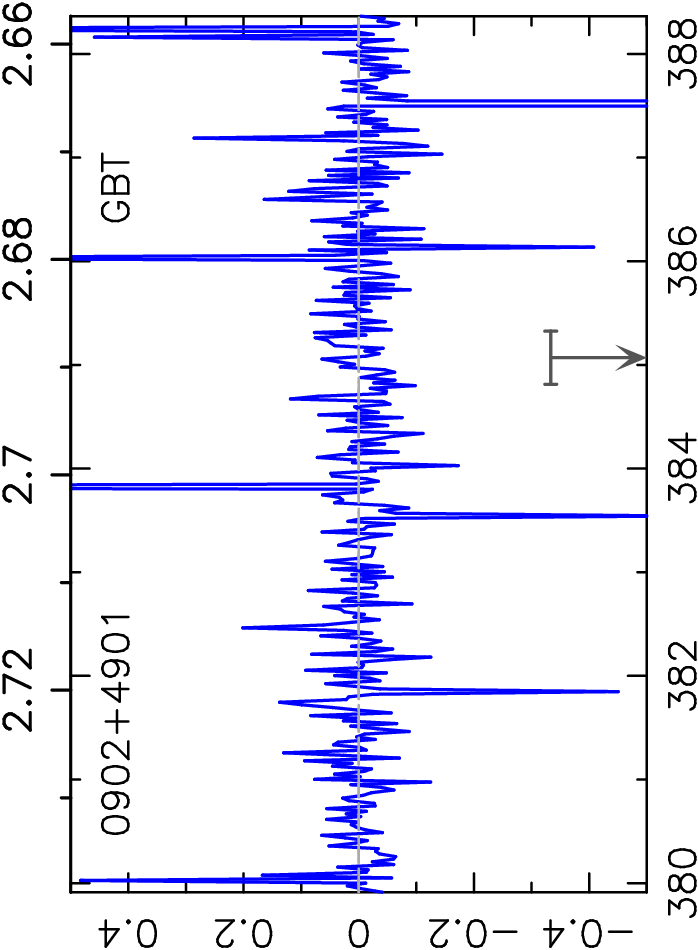}  
\includegraphics{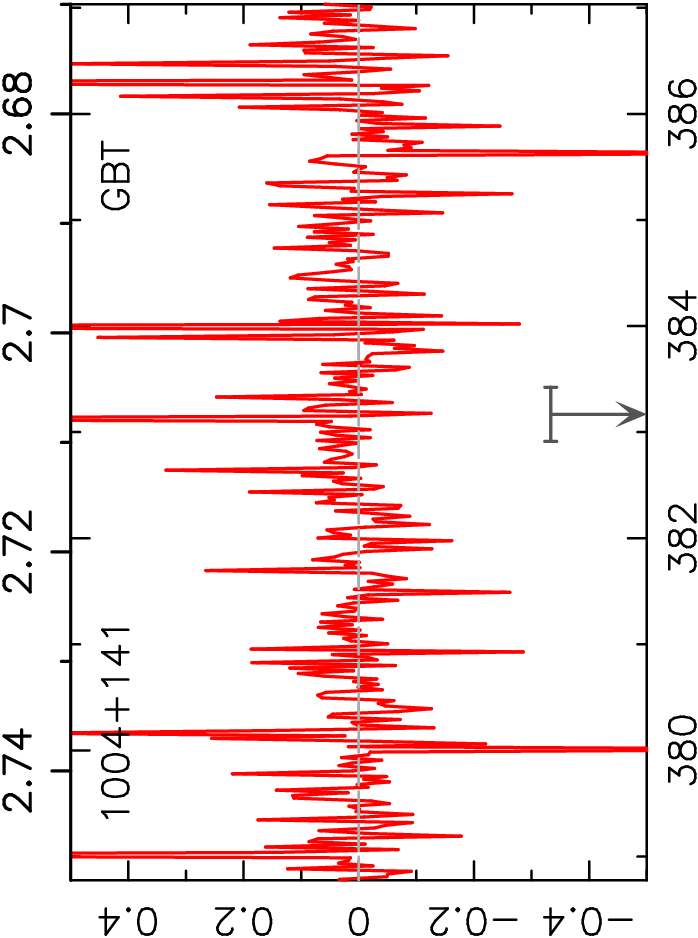}  
\includegraphics{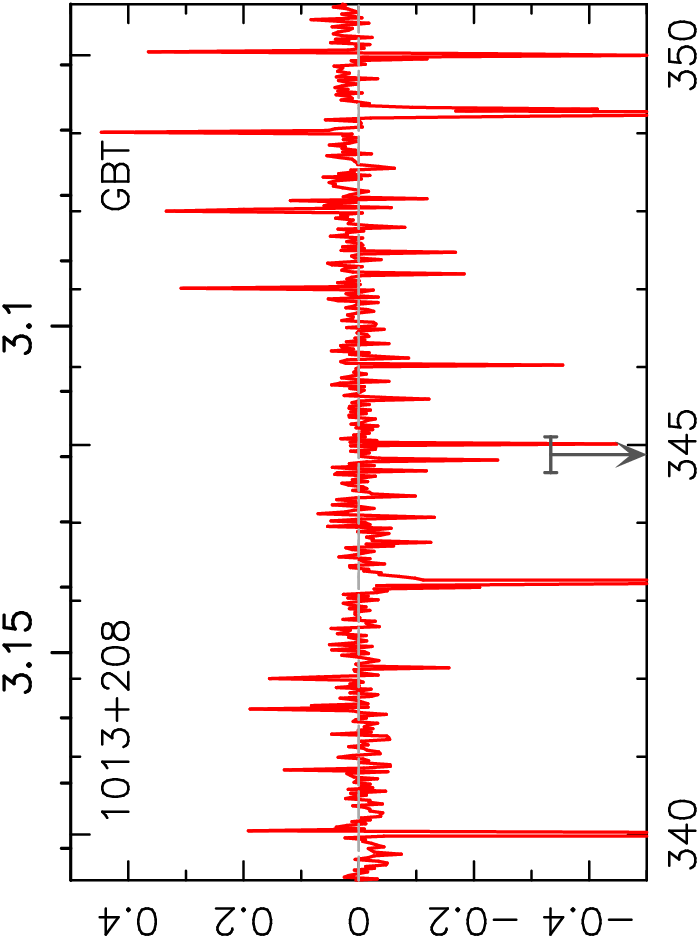} 
\includegraphics{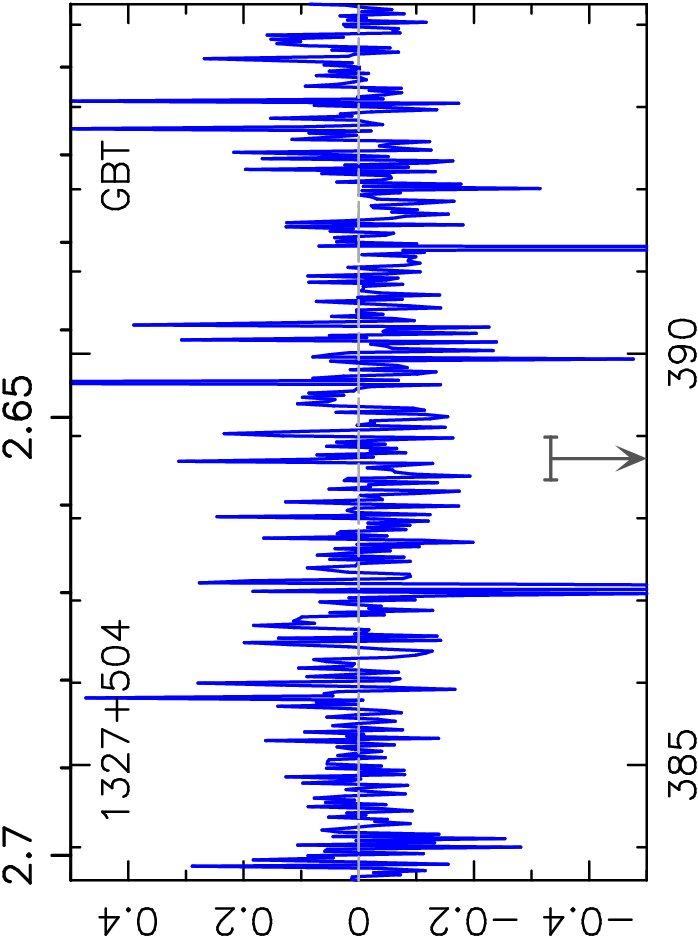} 
\includegraphics{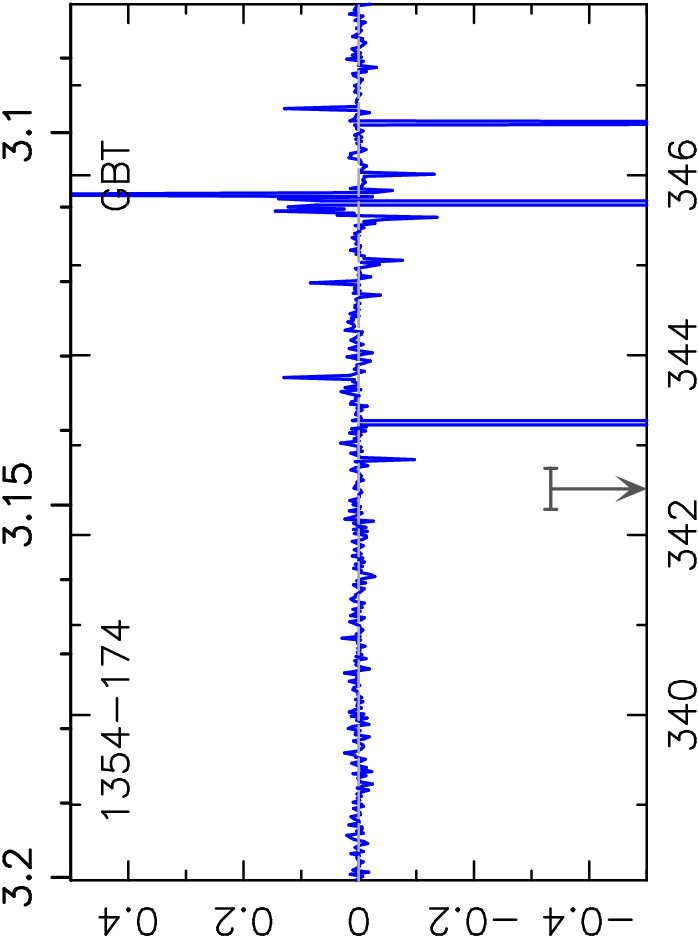}
\includegraphics{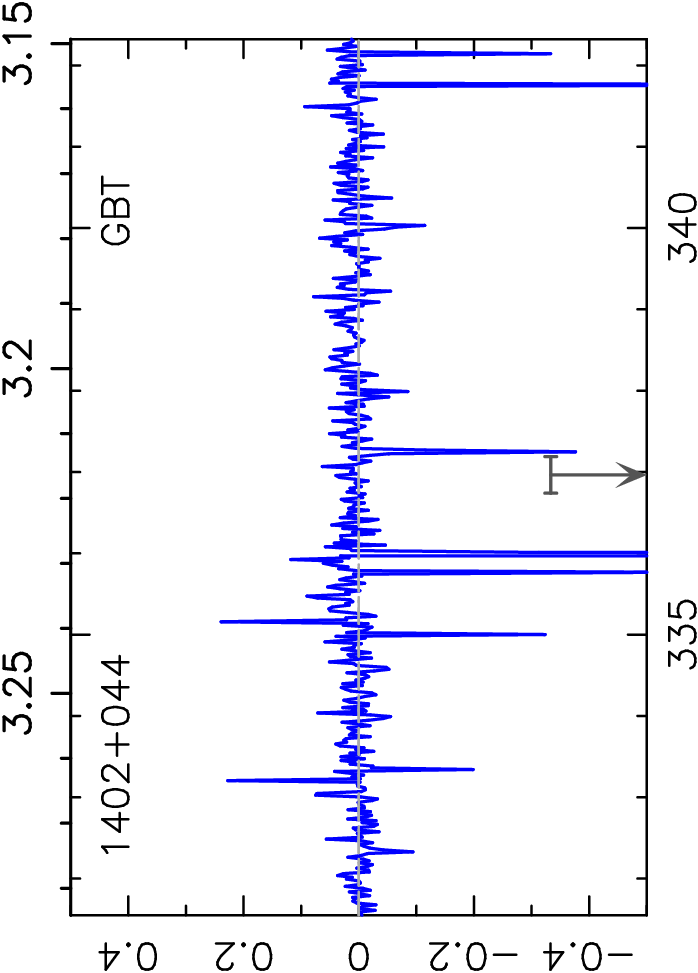}  
\includegraphics{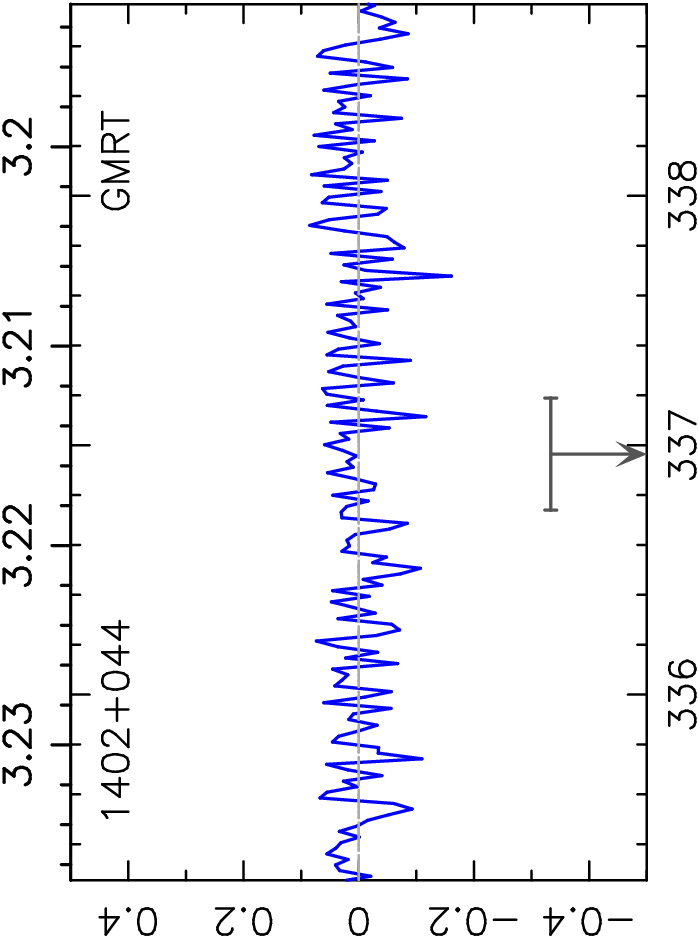} 
\includegraphics{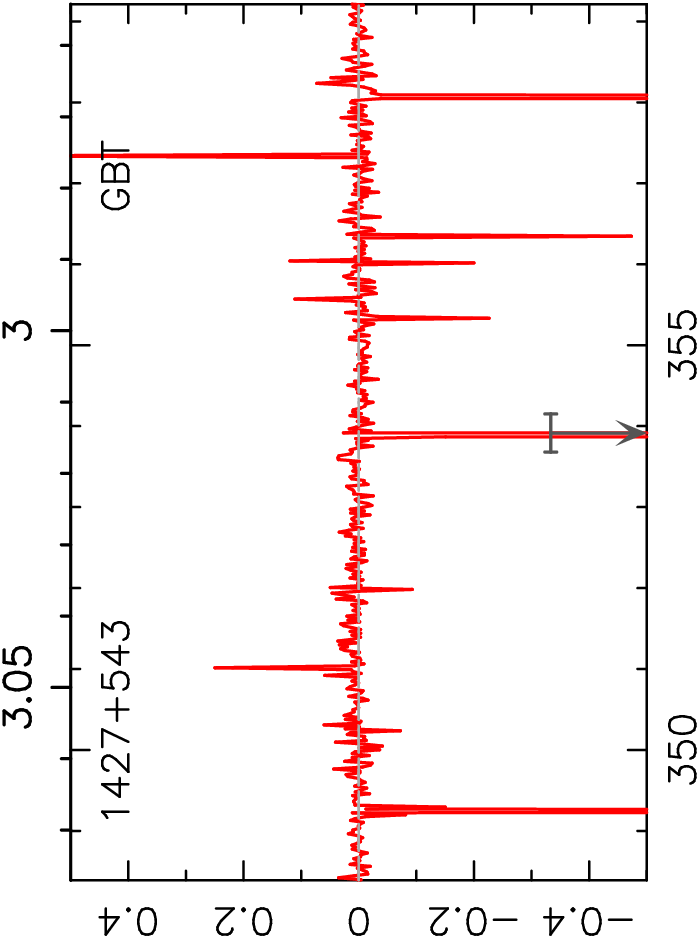} 
\includegraphics{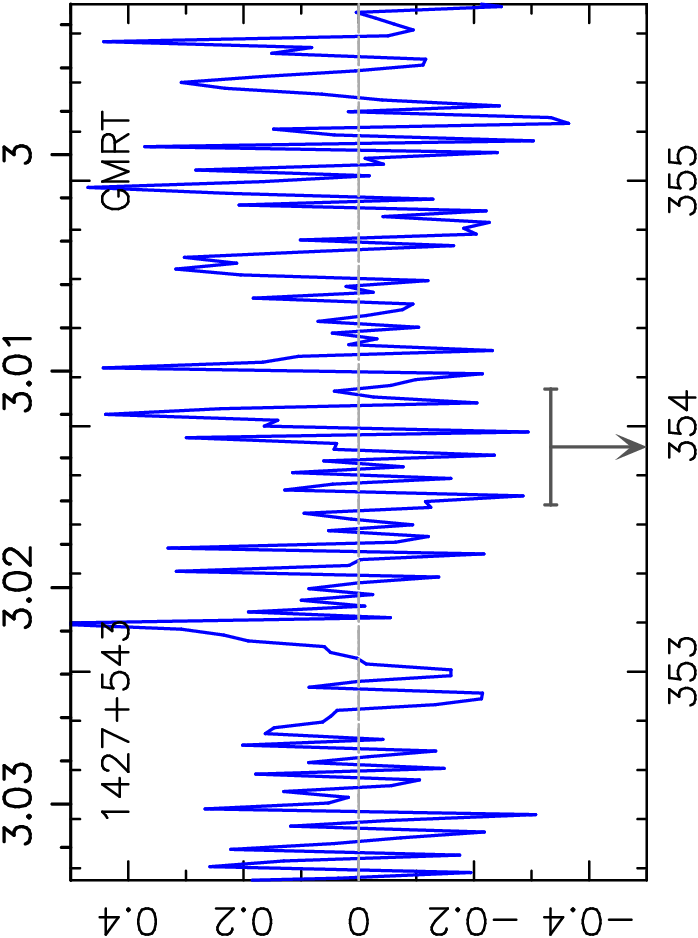}
 %
%\special{psfile= spectra/1446-111-YY.dat-freq_poly2_tau_act_20kms.ps   hoffset=-10 voffset=100 hscale=25 vscale=25 angle=-90}  
\includegraphics{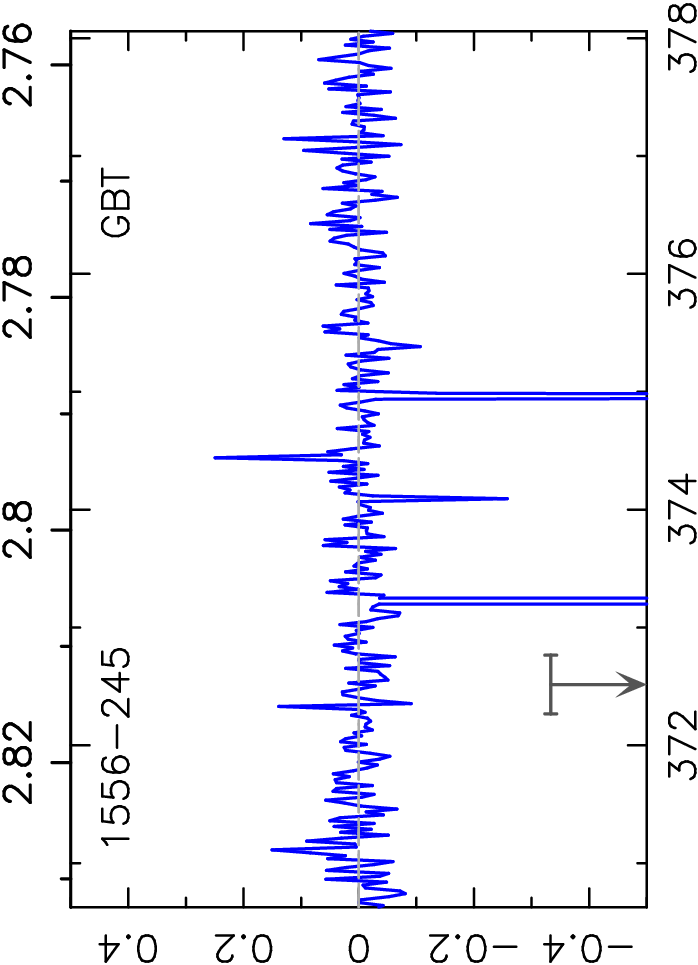}  
\includegraphics{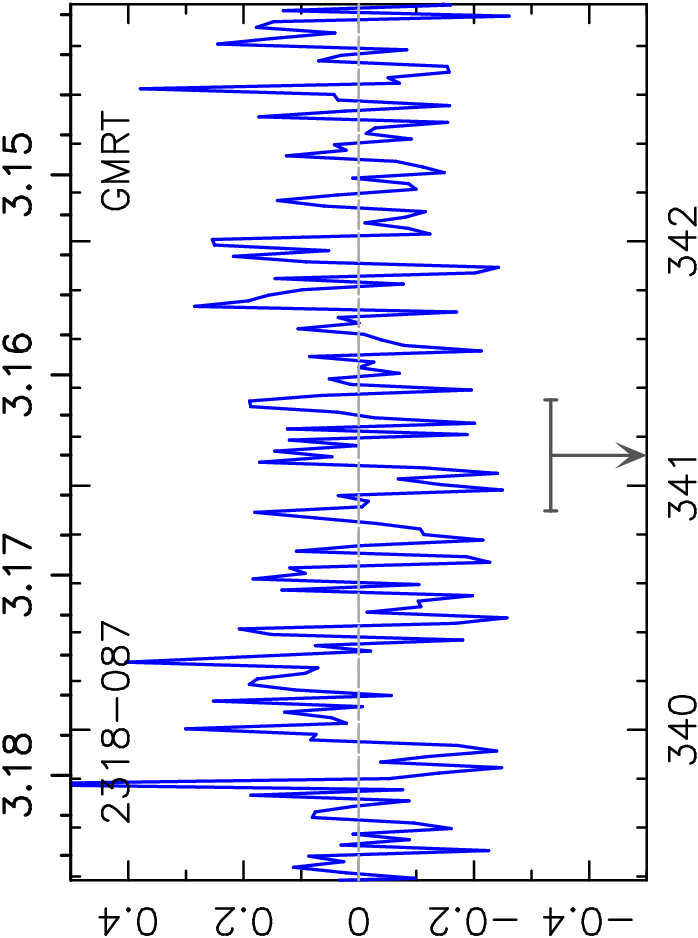} 
\includegraphics{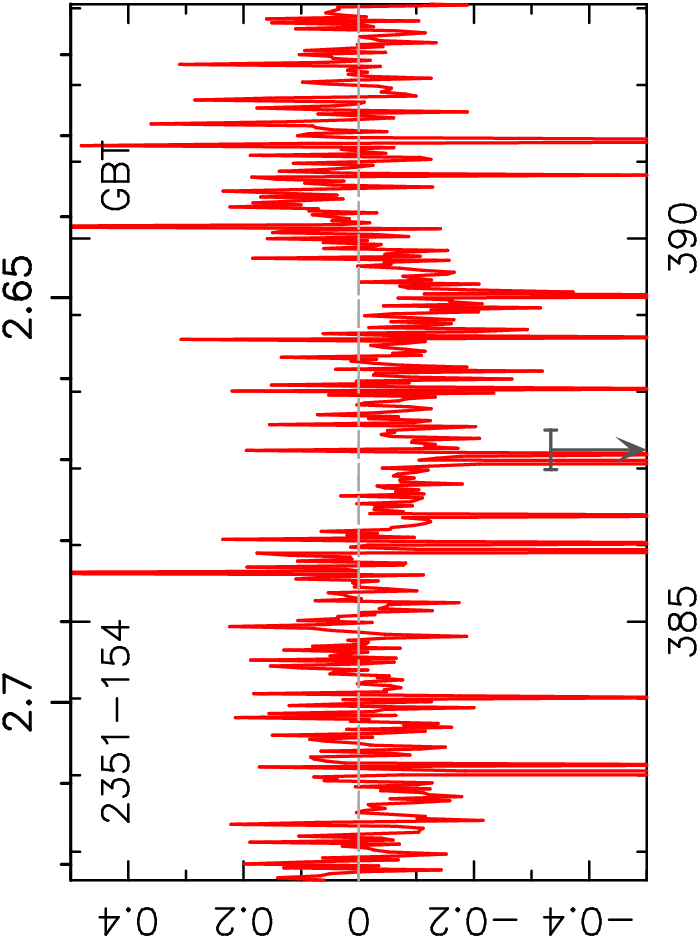} 
\caption{The reduced spectra, for which optical depth limits could be derived (Table \ref{obs}), at a spectral
  resolution of 20 \kms.  The ordinate gives the observed optical depth and the abscissa the barycentric frequency [MHz]. The
  scale along the top of each panel shows the redshift of \HI\ 21-cm over the frequency range and the downwards arrow
  shows the expected frequency of the absorption from the optical redshift, with the horizontal bar showing a span of
  $\pm200$ \kms\ for guidance (the profile widths of the 21-cm detections range from 18 to 475 \kms, with a mean of 167 \kms).
The traces in red have a significant RFI spike within this range, thus potentially masking a detection. These spectra are not included in any further analysis.}
\label{spectra}
\end{figure*}

Of the 29 spectra, three were completely dominated by RFI and a further three had close to zero flux.
In Fig. \ref{spectra} we show the remaining 23 spectra, from 19 different targets.
Of these, seven have RFI/instrumental spikes close to the expected frequency of the putative absorption, which may
conceal any possible detection (Table~\ref{obs}).\footnote{All of the spikes arise in spectra taken with the GBT, since
  the single dish limits the options to mitigate the RFI. Another consequence is that the large beam
 at these frequencies  (${\rm HPBW}\approx40\arcmin$), means that the off-measurement is very likely to contain another
  radio source. This will affect the flux calibration, leading to the differences from the GMRT values for the same source
  (which has ${\rm HPBW}\lapp0.5\arcmin$ and no off-measurement), although the optical depths are expected to be
  unaffected \citep{rie12}.}  This leaves 16 good spectra of 14 different targets, 11 of which have not been previously
searched.
\begin{table*} 
\centering
\begin{minipage}{160mm}
  \caption{The observational results by IAU name (B1950 designation, see Table \ref{NED} for the full names as resolved
    by the NASA/IPAC Extragalactic Database). $z$ is the optical redshift of the source, $\Delta S$ the rms noise
    reached per 20 \kms\ channel, $S_{\rm meas}$ is the measured flux density, $\tau_{3\sigma}$ the derived optical
    depth limit, where $\tau_{3\sigma}=-\ln(1-3\Delta S/S_{\rm meas})$ is quoted for these non-detections. These give the
    quoted column densities, where $T_{\rm s}$ is the spin temperature and $f$ the covering factor;  a blank field
   indicates that the observations were dominated by RFI or the presence of an RFI/instrumental spike close to the expected absorption redshift.
The final  two columns give the frequency and redshift range over which the limit applies (neglecting the spikes).}
\begin{tabular}{@{}l c c c  c l c c  c  @{}} 
\hline
\smallskip
Source     &  $z$  & Tel. & $\Delta S$ [mJy] & $S_{\rm meas}$ [Jy] & $\tau_{3\sigma}$ & $N_{\text{\HI}}$  [\scm]  & $\nu$-range [MHz] & $z$-range \\
\hline
0032+276 & 2.9642 & GBT         & 57 & 1.250 & $<0.14$ & ---& 354.23 -- 359.83 & 2.9474 -- 3.0099 \\  
...               & ...         & GMRT      &38& 0.676   &   $<0.17$  & $<6.1\times10^{18}\,(T_{\rm s}/f)$   &  356.50 -- 360.33  &2.9420 --2.9843 \\
0239+108$^a$  & 2.6800 &  GBT       & 48 & 1.032  &  $<0.14$ &     --- &  380.48 -- 391.06 & 2.6322 -- 2.7332\\ 
0319+121$^b$ & 2.6620 & GBT       & 61 & 1.510 &    $<0.12$  &    --- &  382.51 -- 393.83   & 2.6067 -- 2.7134 \\
0511+009   & 2.6794 & GBT       & 64 &  5.208 &   $<0.04$   &  $<1.3\times10^{18}\,(T_{\rm s}/f)$ & 380.71  -- 390.73   & 2.6352 -- 2.7309 \\
0521--262 & 3.109 & GMRT &   -- & ---     &    ---           &          ----                                     &  \multicolumn{2}{c}{\sc rfi dominant}  \\   
0609+607   & 2.7020 & GBT       & 54  & 1.364 &   $<0.12$   &       $<4.4\times10^{18}\,(T_{\rm s}/f)$  & 378.74 -- 389.48     & 2.6469 -- 2.7503 \\
0620+389$^c$   &  3.4690 & GBT     &  44   & 2.926 &   $<0.05$ &  $<1.6\times10^{18}\,(T_{\rm s}/f)$   & 312.11 -- 323.60 &  3.3894 -- 3.5509 \\
...               & ...         & GMRT      & 11   & 0.226& $<0.14$  &    $<5.0\times10^{18}\,(T_{\rm s}/f)$  & 315.92 -- 319.75 & 3.4422 -- 3.4961\\    
0749+426$^d$  & 3.5892  & GMRT      & 19 & 0.771&    $<0.07$ &  $<2.7\times10^{18}\,(T_{\rm s}/f)$   & 307.60 --    311.43 & 3.5609 -- 3.6177\\
0800+618$^e$   & 3.0330  &  GMRT      & -- & --     &    ---           &          ----                                     &    \multicolumn{2}{c}{\sc rfi dominant}   \\  %  352.192 MHz 
0830+115   & 2.9786 & GBT     & 35 & 3.906&  $<0.04$  &   $<1.3\times10^{18}\,(T_{\rm s}/f)$  &  351.37  -- 359.90 &  2.9467 -- 3.0425 \\
0834--201 & 2.7520& GBT     & 78   & 4.223 &   $<0.06$   &  $<2.0\times10^{18}\,(T_{\rm s}/f)$  &  375.10 -- 384.38    & 2.6953 -- 2.7868 \\
0902+490$^f$    & 2.6887  & GBT    &   60 & 0.916 &  $<0.20$    &  $<7.2\times10^{18}\,(T_{\rm s}/f)$   &  379.26 -- 390.94 & 2.6334 -- 2.7452\\
0913+003   & 3.074  & GMRT  & -- & ---     &    ---           &          ----                                     &     \multicolumn{2}{c}{\sc rfi dominant} \\  % 348.65 MHz
1004+141$^g$     & 2.7070   & GBT   & 110  & 0.614 &   $<0.77^{\dagger}$   &   --- &378.66  -- 386.95  & 2.6708 -- 2.7511\\
1013+208  & 3.1186   & GBT   & 42 &  1.094 &   $<0.12$    &    --- & 340.34  -- 350.54 &  3.0521 -- 3.1735 \\
1327+504  & 2.6540    & GBT   & 69   & 0.682 &   $<0.30$  &    $<1.1\times10^{19}\,(T_{\rm s}/f)$  &383.69  -- 394.47 &  2.6008 -- 2.7020 \\
1354--174  &  3.1470 & GBT   & 41  & 3.912 &   $<0.03$  &    $<1.1\times10^{18}\,(T_{\rm s}/f)$  & 337.49 -- 347.84 & 3.0836 -- 3.2088 \\
1402+044  & 3.2153  & GBT   & 53    & 1.429&  $<0.11$    &  $<4.1\times10^{18}\,(T_{\rm s}/f)$     &   331.44 -- 342.91&  3.1422 -- 3.2856 \\
...               & ...         & GMRT      & 17  & 0.351 &   $<0.15$  &  $<5.3\times10^{18}\,(T_{\rm s}/f)$    & 335.04  -- 338.88 & 3.1915 -- 3.2395 \\  
1427+543  & 3.0134  & GBT   & 47     &  2.606 &  $<0.05$   &   ---  & 348.32 -- 359.74 & 2.9484 -- 3.0779 \\
...               & ...         & GMRT    & 68 & 0.429  &   $<0.65^{\dagger}$  &     $<2.4\times10^{19}\,(T_{\rm s}/f)$    & 352.00 --  355.83 & 2.9918 -- 3.0352\\
1446--111  &  2.6326 & GBT   & 75 & 0.038 & ---           &          ----                                  & 385.43 -- 396.71 &  2.5805 -- 2.6852\\
1556--245  & 2.8130    & GBT   & 65  & 1.754 &  $<0.11$  &   $<4.1\times10^{18}\,(T_{\rm s}/f)$  &   369.89 -- 378.24&  2.7553 -- 2.8401 \\
1614+051  & 3.2150 & GMRT & 11   & 0.024 & ---           &     ---    &335.07 --  338.90 & 3.1912 -- 3.2391 \\   % {\color{red} coords okay} 
1745+624   & 3.8890    & GBT   & 67  &0.045 & ---           &    ---  & 285.20  -- 296.26 &  3.7945 -- 3.9805 \\ % RFI
2318--087   & 3.1639 & GMRT & 74 & 0.505 &  $<0.56^{\dagger}$   &    $<2.0\times10^{19}\,(T_{\rm s}/f)$  &339.21 --  343.04 & 3.1406 -- 3.1874\\  
2351--154  & 2.6680  & GBT   & 81 & 0.794  & $<0.37^{\dagger}$      &   --- & 381.41 --  393.10 &  2.6133 -- 2.7241\\
\hline
\end{tabular}
{Notes: $^a$Observed by \citet{gdb+15} but ruined by RFI, $^b$$N_{\text{\HI}}<6.3\times10^{17}\,(T_{\rm s}/f)$ per 30 \kms\  by  \citet{gdb+15},
 $^c$$N_{\text{\HI}}<3.6\times10^{17}\,(T_{\rm s}/f)$ per 30 \kms\ by \citet{akk16}, $^d$$N_{\text{\HI}}<1.4\times10^{18}\,(T_{\rm s}/f)$ per 30 \kms\  by \citet{akk16}, $^e$$N_{\text{\HI}} <2.1\times10^{18}\,(T_{\rm s}/f)$ per 26 \kms\ by  \citet{akk16}, $^f$$N_{\text{\HI}} <1.5\times10^{18}\,(T_{\rm s}/f)$ per 30 \kms\ by \citet{gdb+15}, 1004+141,  $^g$$N_{\text{\HI}} <1.5\times10^{18}\,(T_{\rm s}/f)$ per 30 \kms\ by  \citet{gdb+15}. $^{\dagger}$Since $\tau\gapp0.3$, the optical depth limit is derived assuming $f=1$ (see \citealt{cwa+17}).}
\label{obs}  
\end{minipage}
\end{table*} 

In the optically thin regime (where $\tau\lapp0.3$), the total neutral hydrogen column density
is related to the velocity integrated optical depth of the \HI\ 21-cm
absorption via
\[
N_{\text \HI}\approx 1.823\times10^{18}\,\frac{T_{\rm s}}{f}\int\!\tau\,dv,
\]
where $T_{\rm s}$ is the spin temperature of the gas, which is a measure of the excitation from the lower hyperfine
level \citep{pf56,fie59}  and $\int\!\tau dv$ is the observed velocity integrated optical depth of the absorption.
In order to compare our limits with previous surveys, all spectra are  re-sampled to the same spectral resolution (20 \kms, as in
Fig. \ref{spectra}), which is used as the FWHM to obtain the integrated optical depth limit, thus giving the $N_{\text
  \HI} f/T_{\rm spin}$ limit per channel (see \citealt{cur12}). 

Of the 11 new  sources searched, for which a limit could be obtained, there were no detections of
\HI\ 21-cm absorption. 
\begin{figure*}
\centering \includegraphics[angle=-90,scale=0.68]{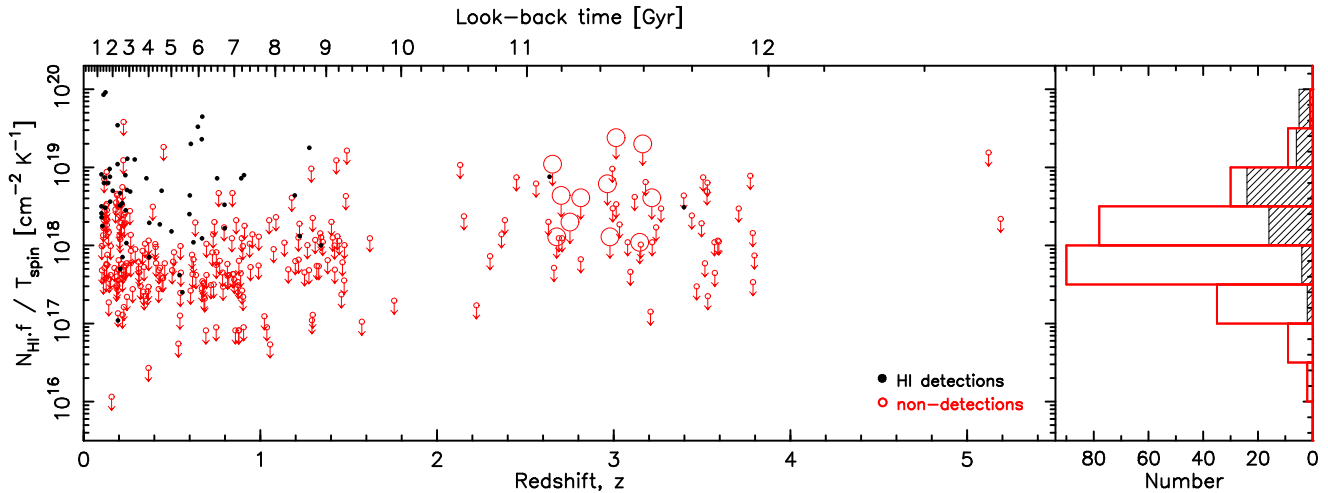}
\caption{The line strength [$1.823\times10^{18}\,(T_{\rm spin}/f) \int\!\tau\,dv$] versus redshift for the $z\geq0.1$
  associated \HI\ 21-cm absorption searches. The filled circles/histogram represent the detections and the unfilled
  circles/histogram the $3\sigma$ upper limits to the non-detections, with the large circles designating our new targets.}
\label{N-z}
\end{figure*}
Although the limits are not as sensitive as many of the other searches, particularly those at low redshift (Fig. \ref{N-z})\footnote{For example, 
at $z>2$ the most sensitive limit is $N_{\text \HI} = 10^{17}\,(T_{\rm s}/f)$ \scm\ and  at $z\leq2$ there are seven searches below this limit.},
there are 42 detections and 116 previous non-detections over the range of sensitivities searched [$N_{\text{\HI}} = 0.1 - 2.4\times10^{19}\,(T_{\rm s}/f)$ \scm]. 
This gives a detection rate of 27\% over all redshifts within these sensitivity limits, which is close to the overall detection
rate (30.5\%,  Sect. \ref{ipr}) and comparable to the 25\% found by \citet{gss+06}. Note that the highest detection rate, of
40\%, was obtained by \citet{vpt+03} in a survey of $z\lapp1$ compact radio sources\footnote{Not only do compact
  sources tend to have lower ultra-violet luminosities \citep{cw10,ace+12}, but there is an anti-correlation between the
  optical depth of the 21-cm absorption and the extent of the radio source \citep{cag+13}.}, where we may expect a
maximum of $\approx50$\%, due to the alignment between the gaseous disk and continuum source along our sight-line (see
\citealt{cw10}). From binomial statistics, we therefore expect $3.8\pm1.6$ new detections for the 14 different targets with good spectra.
 
%If p = 0.3 is the expected detection rate, then the variance in the number of detections in a sample of n objects is var = n*p*(1-p). So your expected standard deviation in the number of detections, assuming a detection rate of 0.3, is sqrt(6*0.3*0.7) = 1.1. SO HERE IS sqrt(14*0.27*0.73) 

\subsection{Ionising photon rates}
\label{ipr}

Following our usual procedure (e.g. \citealt{cwsb12}), for each target  we obtained the photometry from NASA/IPAC Extragalactic Database (NED),
the Wide-Field Infrared Survey Explorer (WISE, \citealt{wem+10}), Two Micron All Sky Survey (2MASS, \citealt{scs+06})
and the Galaxy Evolution Explorer (GALEX data release GR6/7)\footnote{http://galex.stsci.edu/GR6/\#mission} databases. 
Each flux, $S_{\nu} $, was corrected for Galactic extinction \citep{sfd98}, before being converted to a 
specific luminosity  at the source-frame frequency, via $L_{\nu}=4\pi \, D_{\rm L}^2\,S_{\nu}/(z+1)$, 
where $D_{\rm L}$ is the luminosity distance to the source, given by
\[
D_{\rm L} = D(z+1),{\rm ~where~} D = \frac{c}{H_0}\int_{0}^{z}\frac{dz}{H_{\rm z}/H_0} % \text doesn't work for this style
\]
is the line-of-sight co-moving distance (e.g. \citealt{pea99}), 
in which $c$ is the speed of light, $H_0$ the Hubble constant, $H_{\rm z}$ the Hubble parameter at redshift $z$ and 
\[
\frac{H_{\rm z}}{H_{0}} = \sqrt{\Omega_{\rm m}\,(z+1)^3 + (1-\Omega_{\rm m} - \Omega_{\Lambda})\,(z+1)^2 + \Omega_{\Lambda}},
\]
where we use a standard $\Lambda$ cosmology, with $H_{0}=71$~km~s$^{-1}$~Mpc$^{-1}$, $\Omega_{\rm m}=0.27$ and
  $\Omega_{\Lambda}=0.73$.
We then fit a power law to the 
UV rest-frame data, allowing the ionising photon rate, $Q_\text{\HI}\equiv \int^{\infty}_{\nu}({L_{\nu}}/{h\nu})\,d{\nu}$, to be derived from
\[
\int^{\infty}_{\nu}\frac{L_{\nu}}{h\nu}\,d{\nu},~{\rm where}~\log_{10}L_{\nu} = \alpha\log_{10}\nu+ {\cal C} \Rightarrow  L_{\nu} = 10^{\cal C}\nu^{\alpha},
\]
\AL\ is the spectral index and ${\cal C}$ the intercept, which gives
\[
\frac{10^{\cal C}}{h}\int^{\infty}_{\nu}\nu^{\alpha-1}\,d{\nu} = \frac{10^{\cal C}}{\alpha h}\left[\nu^{\alpha}\right]^{\infty}_{\nu} = \frac{-10^{\cal C}}{\alpha h}\nu^{\alpha}~{\rm where}~\alpha < 0,
\]
for the ionising photon rate. 
\begin{figure*}
\centering \includegraphics[angle=-90,scale=0.62]{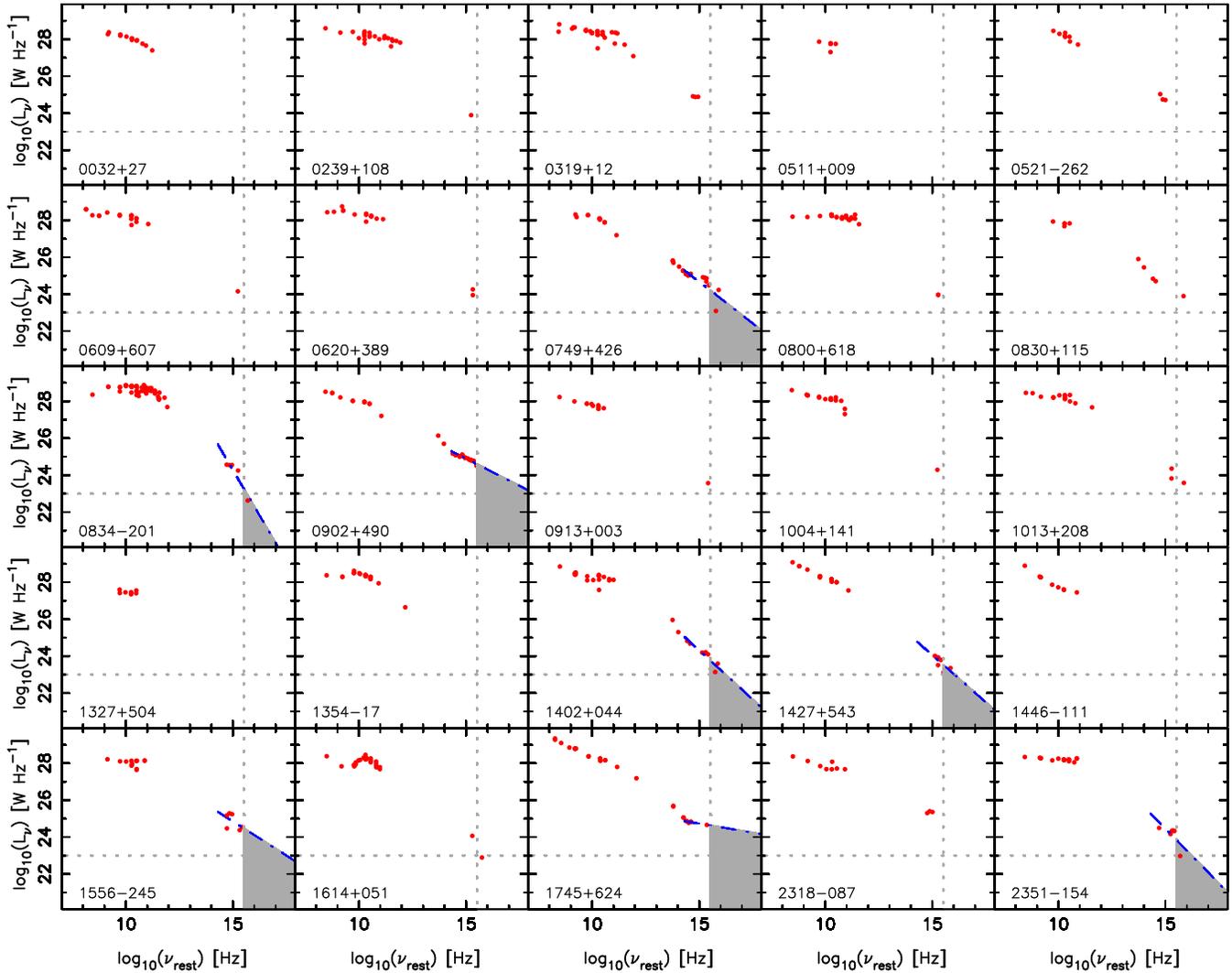}
\caption{The rest-frame photometry for each of the 25 targets. The dashed line shows the power-law fit to the optical/UV data,
the vertical dotted line signifies a rest-frame frequency of $3.29\times10^{15}$ Hz ($\lambda = 912$~\AA) 
and the horizontal line the critical $\lambda = 912$~\AA\  luminosity of $L_{\rm UV}\sim10^{23}$ \WpHz, with the shading
showing the region over which the ionising photon rate is derived (Table \ref{NED}).}
\label{SEDs}
\end{figure*}

\begin{figure*}
\centering \includegraphics[angle=-90,scale=0.68]{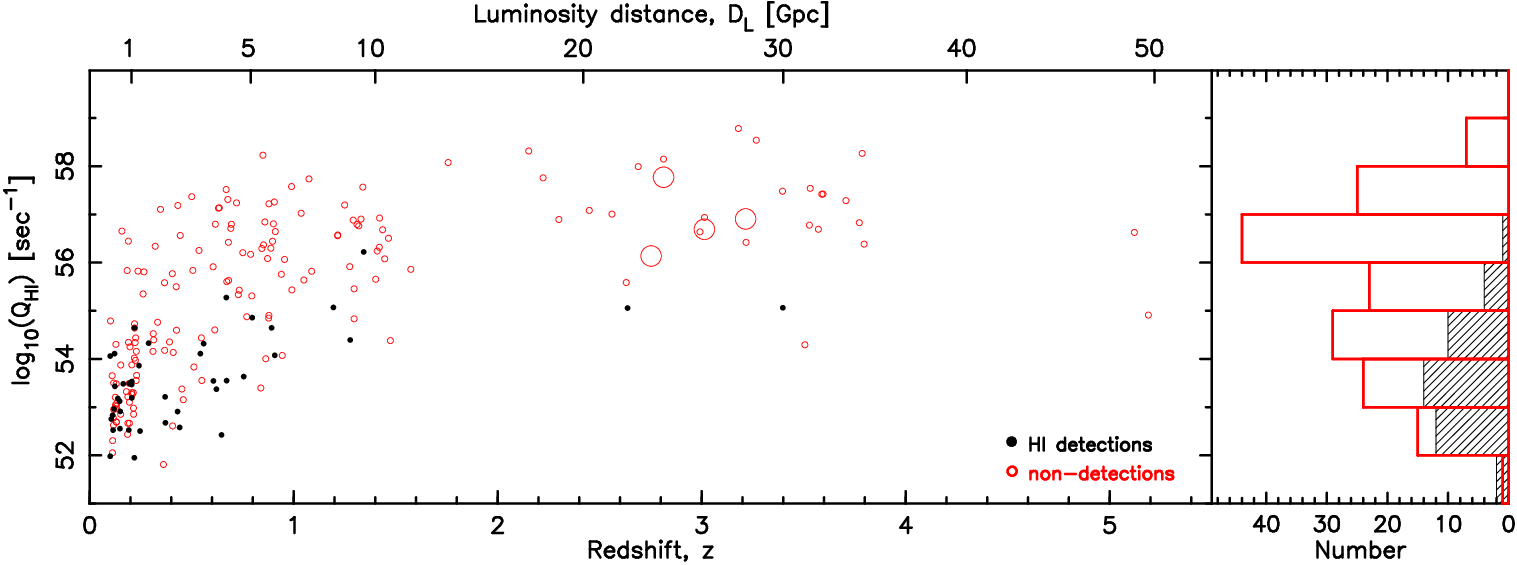}
\caption{The ionising ($\lambda \leq912$ \AA) photon rate versus redshift for the $z\geq0.1$ \HI\ 21-cm absorption
  searches.  The symbols and histogram are as per Fig. \ref{N-z}, where we see that the two $z>2$ detections are below
  the critical ionising photon rate (4C\, +05.19 at $z=2.64$, \citealt{mcm98} \& B2\,0902+34 at $z=3.40$,
  \citealt{ubc91}). Note that 0749+426 at $z =3.5892$--$\log_{10}Q_\text{\HI} = 57.71$ is not flagged as one of our targets since
  this was observed to a deeper limit by \citet{akk16}, Table \ref{obs}.}
\label{Q-z}
\end{figure*}

Although there was only sufficient photometry for four of the targets which also had good spectra (Fig. \ref{SEDs}), we
can add these to the previous searches (Fig. \ref{Q-z})\footnote{Compiled from
  \citet{dsm85,mir89,vke+89,ubc91,cps92,cmr+98,cwh+07,mcm98,ptc99,ptf+00,rdpm99,mot+01,ida03,vpt+03,cwm+06,cww+08,cwm+10,cwwa11,cwsb12,cwt+12,caw+16,cwa+17,gss+06,omd06,kpec09,ems+10,ssm+10,css11,cgs13,ace+12,asm+15,ysdh12,ysd+16,gmmo14,sgmv15,akk16,akp+17,gdb+15}.}.
%      awk < all-info.dat '{if ($3 >= 0.1 && $4 > 0) print $0}' | wc -l  311    awk < all-info.dat '{if ($3 >= 0.1 && $4 > 0 && $24 > 0) print $0}' | wc -l  211
Of the 311 $z\geq0.1$ sources for which limits can be obtained, 211 have sufficient photometry to determine the ionising photon rate. Of these, there are
43 detections and 98 non-detections with ionising photon rates up to the highest value where 21-cm absorption has been detected
($Q_\text{\HI}\leq1.7\times10^{56}$~s$^{-1}$, \citealt{kpec09}). Applying this 30.5\% detection rate to the
$Q_\text{\HI}> 1.7\times10^{56}$~s$^{-1}$ sources, gives a binomial probability of $8.72\times10^{-12}$ of obtaining 0
detections and 70 non-detections, which is significant at $6.83\sigma$, assuming Gaussian
statistics. This strengthens the case for a critical UV luminosity hindering the detection of \HI\ 21-cm absorption at high redshift.
%includes \citet{gdb+15}
\begin{table} 
\caption{The ionising photon rates [photons s$^{-1}$] of the targets for which a limit to the absorption strength could be obtained.}
\begin{tabular}{@{}l c c c   @{}} 
\hline
\smallskip
NED  name&  IAU & $z$  &  $\log_{10}Q_{\text{\HI}}$ \\
\hline
B2\,0032+27 & 0032+276 & 2.9642 & --- \\ % ?   00:34:43.48 +27:54:25.72 358.418 MHz  1.07335  Jy, 94 - S(1.4GHz) = 1.089+/0.0315 Jy at 358.31 MHz
PKS\,0239+108 & 0239+108  & 2.680  & ---\\ % QSO  02:42:29.17 +11:01:00.72 385.976 MHz  1.60251  Jy, 135  - crashes, but no UV NED - flux 1.4 Jy at 365 MHz
PKS\,0319+12 & 0319+121   & 2.662  &   ---\\ % QSO     03:21:53.10 +12:21:13.95   387.873 MHz  2.22895  Jy, 146 -  S(1.4GHz) = 2.188+/0.1833 Jy at 387.88 MHz
PMN\,J0513+0100  & 0511+009   & 2.6770 &  --- \\ %  ?  05:13:40.03 +01:00:21.65  386.296 MHz  3.68774  Jy, 177 -  S(1.4GHz) = 3.695+/0.1615 Jy at 386.29 MHz 
PKS\,0521--262$^{\dagger}$  & 0521--262 & 3.109  & --- \\ % 05:23:18.46 -26:14:09.55 80.82696  -26.23599   345.685 MHz  1.52346  Jy 
BZQ\,J0614+6046 & 0609+607   & 2.702 & --- \\%  QSO  06:14:23.86 +60:46:21.75 383.69 MHz  1.30293  Jy, 195 - crashes, but no UV NED 
B2\,0620+38 & 0620+389   &  3.469 &  ---\\% 06:24:19.02 +38:56:48.73 317.834 MHz  1.25176, 198 - S(1.4GHz) = 1.301+/0.1211 Jy at 317.84 MHz
B3\,0749+426  & 0749+426  & 3.5892   & $57.71\pm0.78$\\  % 07:53:03.33 +42:31:30.76 309.514 MHz  0.578549  Jy, 227 -S(1.4GHz) = 0.599+/0.0389 Jy at 309.15 MHz 
WISE\,J080518.15+614424.0$^{\dagger}$& 0800+618   & 3.033    & ---\\  % 08:05:18.17 +61:44:23.70  352.192 MHz  0.875125  Jy, 234 -  S(1.4GHz) = 0.861+/0.0564 Jy at 352.20 MHz
\protect[HB89]\,0830+115   & 0830+115   & 2.9786  & ---\\  % 08:33:14.36 +11:23:36.23   357.01 MHz  1.90928  Jy, 250 -  S(1.4GHz) = 1.913+/0.0529 Jy at 357.01 MHz
\protect[HB89]\,0834--201 & 0834--201  & 2.752 &  $57.52\pm0.12$\\ % 08:36:39.21 -20:16:59.50  378.573 MHz  2.70527  Jy, 254 - S(1.4GHz) = 2.757+/0.1451 Jy at 378.57 MHz
SDSS\,J090527.46+485049.9$^{\dagger}$ & 0902+490   & 2.6887  &  $57.99\pm0.20$ \\ % 09:05:27.46 +48:50:49.96 136.36444  48.84721   385.07 MHz  1.31202  Jy 
SDSS\,J091551.69+000713.2$^{\dagger}$  & 0913+003    & 3.074  & ---\\ % 09:15:51.69 +00:07:13.30 138.96539  0.12036   348.65 MHz  0.521471  Jy 
 \protect[HB89]\,1004+141 & 1004+141   & 2.707  & ---\\ % 10:07:41.49 +13:56:29.60 383.169 MHz  1.60616  Jy, 291 - crashes but no UV
 PKS\,1014+208& 1013+208   & 3.1186 & ---\\ % 10:16:44.32 +20:37:47.30   344.874 MHz  1.14562  Jy, 294 -  S(1.4GHz) = 1.114+/0.1155 Jy at 344.88 MHz 
SDSS\,J132905.80+500926.5 & 1327+504   & 2.654 & ---\\ % 13:29:05.80 +50:09:26.40 388.723 MHz  1.77807  Jy, 349 - S(1.4GHz) = 0.637+/0.0644 Jy at 388.73 MHz 
PKS\,1354--17  & 1354--174  &  3.147 & ---\\ % 13:57:06.07 -17:44:01.90  342.515 MHz  1.34924  Jy, 357 - S(1.4GHz) = 1.349+/0.0848 Jy at 342.51 MHz T
\protect[HB89]\,1402+044 & 1402+044   & 3.2153  &  $57.04\pm0.47$\\% 14:05:01.11 +04:15:35.81  336.961 MHz  1.38134  Jy, 362 -  S(1.4GHz) = 1.374+/0.1629 Jy at 336.96 MHz
SDSS\,J142921.87+540611.1& 1427+543   & 3.0134  & $56.69\pm0.51$\\ % 14:29:21.87 +54:06:11.12 353.916 MHz  2.41813  Jy , 371 - S(1.4GHz) = 2.196+/0.0430 Jy at 353.92 MHz 
PKS\,B1446-111$^{\dagger}$ & 1446--111  &  2.6326 & ---\\ %14:48:51.16 -11:22:15.73 222.21317  -11.37104   391.084 MHz  1.14506  Jy 
\protect[HB89]\,1556--245 & 1556--245  & 2.813   & $57.77\pm0.78$ \\ % 15:59:41.40 -24:42:38.83   372.52 MHz  1.01608  Jy, 399 - S(1.4GHz) = 1.017+/0.1563 Jy at 372.52 MHz
\protect[HB89]\,1614+051$^{\dagger}$     & 1614+051    & 3.2150  & ---\\ % 16:16:37.55 +04:59:32.73 244.15649  4.99243   336.985 MHz  0.634337  Jy 
4C\,+62.29$^{\dagger}$ & 1745+624    & 3.889   &  $58.51\pm0.07$ \\  % 17:46:14.03 +62:26:54.73 266.55847  62.44854   290.529 MHz  2.04094  Jy 
PKS\,2318--087 & 2318--087   & 3.1639   & ---\\ % 23:21:18.25 -08:27:21.52 341.122 MHz  0.558753  Jy, 485 -  S(1.4GHz) = 0.559+/0.1275 Jy at 341.12 MHz 
\protect[HB89]\,2351--154   &   2351--154     & 2.668 &  $56.94\pm0.90$\\  % 23:54:30.19 -15:13:11.21 387.24 MHz  1.09089  Jy, 495 - S(1.4GHz) = 1.108+/0.0521 Jy at 387.24 MHz
\hline
\end{tabular}
{$^{\dagger}$Shown for completeness as no sensitivity limit could be derived.}
\label{NED}
\end{table}

\section{Summary}

We have undertaken a survey for associated \HI\ 21-cm absorption at redshifts of $z=2.6-3.9$ in 25 radio sources with
the Green Bank and Giant Metrewave Radio Telescope, from which we obtain zero detections. Of the 14 for which limits
could be obtained, 11 are new, with ten of these reaching optical depths of $\tau_{3\sigma}\leq0.3$.
All of the targets were selected from the sources in the {\em Second Realization of the
  International Celestial Reference Frame by Very Long Baseline Interferometry}, for which 1682 have measured redshifts
spanning up to $z =6.2$.  Our intent was to undertake a large survey over all of this redshift space in order to
quantify the incidence of associated \HI\ 21-cm absorption over the history of the Universe. However, observing time was
only awarded for the highest redshift sample, where we do not expect absorption due to all of the neutral gas being
ionised in the $z\gapp1$ sources that have a reliable optical redshift \citep{cw12}.

Nevertheless, the ICRF2 provides a large new sample of strong flat-spectrum radio sources within which to search for
21-cm absorption, and the expectation of a null result should not prevent us from testing it. Given the general
$\approx30$\% detection rate at the sensitivities obtained, we expect $\approx4$ new detections, although obtaining zero
detections is within $\approx2\sigma$ of this. In the context of all the previous searches, there still remains only two
detections in this redshift range \citep{ubc91,mcm98}, both of which are below the ionising photon rate of
$Q_\text{\HI}\sim10^{56}$~s$^{-1}$ (a monochromatic $\lambda =912$ \AA\ luminosity of $L_{\rm UV}\sim10^{23}$
\WpHz).  The addition of the five new sources for which there is sufficient photometry\footnote{The four new plus one
  searched in common with \citet{akk16}.}, gives a binomial probability of $8.72\times10^{-12}$ of the observed
distribution occuring by chance.  This is significant at $6.83\sigma$, which supports the hypothesis that $Q_\text{\HI}\sim10^{56}$~s$^{-1}$
is sufficient to ionise all of the neutral gas in a large spiral galaxy \citep{cw12}.  This appears to be  
 an ubiquitous effect, independent of source selection and applicable over all
redshifts, with the selection of sources with $Q_\text{\HI}\lapp10^{56}$~s$^{-1}$ being required to detect
neutral hydrogen.  This ionising photon rate corresponds to blue magnitudes of $B\gapp22$ at $z\gapp3$, for which there are
currently no radio sources with a measured optical redshift known \citep{cwt+12,msc+15}. 

We therefore reiterate that the detection of neutral gas within the hosts of high redshift active galaxies requires us
to dispense with the usual reliance upon an optical redshift to which to define the search frequency. The way
forward is through wide-band observations of radio sources of unknown redshift  with the Square Kilometre Array.

\section*{Acknowledgements}
We wish to thank the anonymous referee for their helpful comments, as well as Katie Grasha and Jeremy Darling for a
draft manuscript of their forthcoming paper.
We also thank the staff of the GMRT who have made these observations possible. GMRT is run by the National Centre for Radio
Astrophysics of the Tata Institute of Fundamental Research.  The National Radio Astronomy Observatory is a facility of
the National Science Foundation operated under cooperative agreement by Associated Universities, Inc This research has
made use of the NASA/IPAC Extragalactic Database (NED) which is operated by the Jet Propulsion Laboratory, California
Institute of Technology, under contract with the National Aeronautics and Space Administration. This research has also
made use of NASA's Astrophysics Data System Bibliographic Services.

%\bibliographystyle{../mn2e}  
%\bibliography{aa,ref}

\label{lastpage}

\end{document}